\documentstyle[twoside,fleqn,espcrc2,epsf]{article}

\newcommand{\AmS}{{\protect\the\textfont2
  A\kern-.1667em\lower.5ex\hbox{M}\kern-.125emS}}

\title{Topological structure of the SU(3) vacuum and exceptional
eigenmodes of the improved Wilson-Dirac operator\thanks{presented
by Douglas Smith}}

\author{D.Smith\address{Dept of Physics and Astronomy,
        University of Edinburgh, 
        Edinburgh, Scotland}, %
        H.Simma\address{DESY-Zeuthen, Platanenallee 6, 
	D-15738 Zeuthen,Germany}, %
        and 
	M.Teper\address{Dept of Physics,
	University of Oxford, Oxford, U.K.} %
        ({\it UKQCD Collaboration})}
       
\begin{document}

\begin{abstract}
 We present a study of the instanton size and spatial distributions in
 pure SU(3) gauge theory using under-relaxed cooling. 
 We also investigate the low-lying eigenmodes of the (improved)
 Wilson-Dirac operator, in particular, the appearance of zero-modes and
 their space-time localisation with respect to instantons in the 
 underlying gauge field.
\end{abstract}

\maketitle

\section*{Instanton content of the SU(3) vacuum \cite{teper}}

The importance of the instanton content of $SU(3)$ gauge theory 
comes through both the intrinsic importance of
understanding the ground state of QCD and the role instantons
are conjectured to play in light hadron structure. Cooling
is a technique for removing the high-frequency non-topological excitations
of the gauge-field. However, during cooling instantons are also
removed; either if they are very small (lattice artifacts) or through 
$I \bar{I}$ annihilation. We use under-relaxed cooling to
reduce the latter problem.
Also, on the cooled configurations there is still the problem
of extracting the instanton properties; for this we have developed
pattern-recognition algorithms.
We present results for 20 configurations at $\beta=6.0$ $(16^348)$ 
and $\beta=6.2$ $(24^348)$ lattices.

The gauge update for under-relaxed cooling \cite{michael} is
implemented in each Cabibbo-Marinari subgroup as
\begin{equation}
U_{new} = c (U_{min} + \alpha U_{old})
\end{equation}
where $U_{min}$ is the gauge link that minimises the action, $U_{old}$
the original link, $\alpha$ is the under-relaxation parameter 
and $c$  a normalisation constant.
Under-relaxed cooling increases the number of {\it calibrated} sweeps
needed to annihilate an $I \bar{I}$ pair; for a given value of 
$\alpha$ a  calibrated sweep is the number of sweeps needed to destroy
a $\rho = 2$ instanton.
With no under-relaxation one occasionally finds a very narrow
instanton broadening out under cooling (presumably because of its
environment). We have not observed this with (significant)
under-relaxation. We chose $\alpha = 1$ and
our measurements were carried out between 23 and 46 cooling sweeps 
(corresponding to between 10 and 20 cooling sweeps at $\alpha=0$).

On the cooled configurations we first find all the local extrema  of the
symmetrised 
topological charge density, $Q(x)$, relative to the $3^4$ block
surrounding each point. (We do not consider the action, $S(x)$, as it 
clearly records less structure.)
Each peak is treated as a linear
superposition of the topological charge of the object at that point, 
calculated from a lattice-corrected formula,
plus a contribution from every other object on the lattice, calculated
from the continuum formula. A self-consistent set 
of widths is then found by iteration. These are our candidate instantons.


Summing up $Q(x)$ over the lattice and comparing it
to $n_I - n_{\bar I}$ shows a discrepancy. We define
\begin{equation}
\delta = < |Q - (n_I - n_{\bar I})| >
\end{equation}
and impose filters on our candidate instantons. 
The parameters of the filters are chosen to minimise
$\delta$. We have a ``spatial'' filter to remove spurious peaks due 
to ripples on large objects and a ``width'' filter. The latter
compares the width calculated above with the width calculated from
the charge within a radius 2 (or 3) of the peak using a 
lattice-corrected formula; a peak is only included if
the various widths are in sufficiently good agreement.
Full details will appear elsewhere~\cite{teper}.

In Figure 1 we show the instanton size distribution for $\beta=6.2$
at 23 sweeps.
\begin{figure}[htb] \label{figure1}  
\centerline{ \epsfysize=6.0cm 
             \epsfxsize=6.0cm 
             \epsfbox{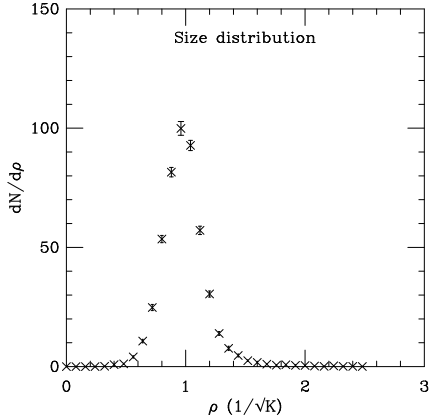}}
\vspace{-10mm}
\caption{Size distribution: $\beta=6.2$, 23 sweeps}
\end{figure}
The distribution is peaked around $\rho \approx
\frac{1}{\sqrt K}$. The best fit to the large-$\rho$ 
tail of the distribution
is $D(\rho) \propto \rho^{-\alpha}$ with $\alpha \approx 10$.
As one would expect, the total number of instantons is found to vary
rapidly with the amount of cooling. However the average size
and the form of the small/large $\rho$ tails varies much less.
We note that our results are consistent
with those of \cite{zurich} but not with those of \cite{boulder}. 

The fact that the impact of a cooling sweep does not scale
complicates the scaling analysis. Figure~2 shows
that we can tune the number of cooling sweeps so as to
get scaling when comparing $\beta=6.2$ and $\beta=6.0$.
\begin{figure} \label{figure2}  
\centerline{ \epsfysize=6.0cm 
             \epsfxsize=6.0cm 
             \epsfbox{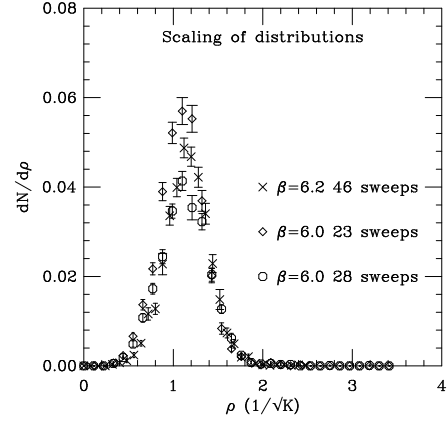}}
\vspace{-10mm}
\caption{Scaling of size distributions}
\end{figure}

Examining the number of unlike charges a distance $R$ away from each
peak (normalised by the volume of the shell) gives the distribution 
that is shown in Fig 3 (for $\beta=6.2$ after 23 sweeps). 
\begin{figure}[htb] \label{figure3}  
\centerline{ \epsfysize=6.0cm 
             \epsfxsize=6.0cm 
             \epsfbox{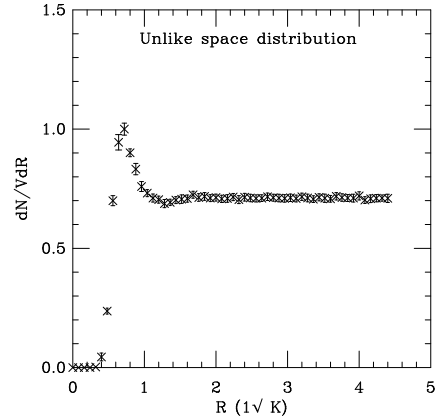}}
\vspace{-10mm}
\caption{ Unlike Spatial Distribution}
\end{figure}
It is uniform at long distances and amplified at short distances.
The corresponding distribution
for like charges is uniform at long distances and suppressed at short
distances. This
implies some screening of instantons by anti-instantons
in the vacuum -- as expected.  


Calculating $<\frac{Q}{|Q|}\frac{q(\rho)}{n(\rho)}>$ where
$n(\rho)$ is the number of objects of size $\rho$ and $q(\rho)$ is the
charge carried by objects of size $\rho$ shows that the charge
carried by small (large) instantons is correlated (anti-correlated)
with the sign of $Q$ (Figure ~4). Indeed our results
suggest over-screening of large instantons by small anti-instantons.

\begin{figure}[htb] \label{figure4}  
\centerline{ \epsfysize=6.0cm 
             \epsfxsize=6.0cm 
             \epsfbox{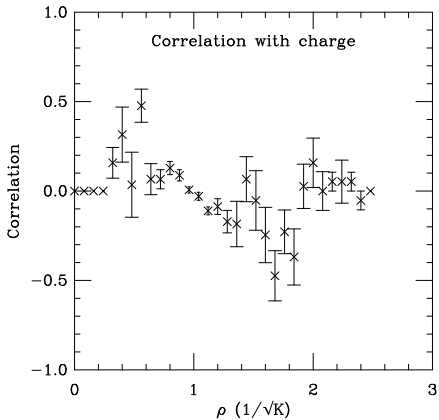}}
\vspace{-10mm}
\caption{$<\frac{Q}{|Q|}.\frac{q(\rho)}{n(\rho)}>$, $\beta=6.2$, 23 sweeps }
\end{figure}
\section*{Zero modes of the improved Wilson-Dirac Operator \cite{simma}}
The Wilson-Dirac operator, or equivalently ${\bf Q}(\kappa)=\gamma_5M$, may
have vanishing or almost-zero eigenvalues for certain values of $\kappa$. 
Moreover, it has been shown that the lowest eigenvalues of ${\bf Q}$ are
strongly localised in space-time \cite{jansen}. 

``Exceptional'' configurations are assumed to be related to
the appearance of (almost-)zero eigenvalues at some $\kappa_0 < \kappa_{crit}$.
They seem to be more frequent at smaller $\beta$ and with SW improvement.
On the other hand, it is unclear if and how these zero modes are related to 
the topology of the underlying gauge field, in particular because the chirality
$\chi = \langle \psi\vert\gamma_5\vert\psi\rangle$ of the corresponding
eigenvectors $\psi$ of ${\bf Q}(\kappa_0)$ is typically much smaller than one.
To clarify this relation we investigate the low-lying eigenmodes of
${\bf Q}(\kappa)$
using a modified conjugate gradient method~\cite{cg}. 
We verified that the index theorem is realized on single instanton
configurations (generated as in ref.~\cite{smit}), for both
improved and unimproved Wilson Fermions, with one
right-handed (4 right-handed plus 3 left-handed) zero modes in the range
$\kappa < 0.2$ for anti-periodic (periodic) boundary conditions. This is
in complete analogy to the results for 2-dimensional Wilson and for staggered 
fermions \cite{smit}.

\begin{table}
\caption{Position $\kappa_0$ and chirality $\chi$ of the zero-mode
for single instantons without (with) improvement.}
\begin{center}
\begin{tabular}{|c|c|c|c|}\hline
$\rho$ & $c_{sw}$ & $\kappa_0$ & $\chi$ \\ \hline
   2    &   0 (1)  &   0.135 (0.126)  &   0.66 (0.986) \\
   3    &   0 (1)  &   0.129 (0.125)  &   0.79 (0.999) \\
   4    &   0 (1)  &   0.127 (0.125)  &   0.90 (0.999) \\ \hline
\end{tabular}
\end{center}
\end{table}

For anti-periodic boundary conditions, the eigenmodes are localised and 
centered on the instanton. Moreover, we find considerable effects from
discretization errors and SW improvement which are summarized in table~1.

We also investigated the localisation of the lowest eigenmodes on 
four exceptional configurations encountered by UKQCD at $\beta=6.0$ on
$16^348$ and $32^364$ lattices. In all cases the (almost-)zero mode is localized
close to a small ($\rho=2a \ldots 3a$) instanton (less than $\sqrt{2}a$ away).
The chirality has the same sign as the topological charge of the topological
object but is not correlated with the overall topological charge of the 
configuration.
This provides some evidence that exceptional configurations are related 
to small instantons in the underlying gauge field.
More detailed results will be presented elsewhere \cite{simma}.

\section{Acknowledgements}
We acknowledge financial support by PPARC grant GR/K41663.
D.S. was funded by the Carnegie Trust for the Universities of Scotland.

\end{document}